\documentclass[journal,twoside,web]{ieeecolor}
\usepackage{generic}
\usepackage{cite}
\usepackage{amsmath,amssymb,amsfonts}
\usepackage{graphicx}
\usepackage{textcomp}

\usepackage{multicol}
\usepackage{ulem}
\usepackage{url}
\usepackage{hyperref}
\usepackage{multirow}
\usepackage{adjustbox}
\usepackage[font=small]{caption}

\usepackage[linesnumbered,ruled,vlined]{algorithm2e}
\SetKwInput{KwInput}{Input}                
\SetKwInput{KwOutput}{Output}              

\newcommand{\orcid}[1]{\href{https://orcid.org/#1}{\includegraphics[width=10pt]{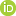}}}

\def\BibTeX{{\rm B\kern-.05em{\sc i\kern-.025em b}\kern-.08em
    T\kern-.1667em\lower.7ex\hbox{E}\kern-.125emX}}
\begin{document}
\title{A network-based analysis of disease modules from a taxonomic perspective}
\author{Giorgio Grani \orcid{0000-0002-0388-1283}, Lorenzo Madeddu \orcid{0000-0002-5702-8691},  and Paola Velardi \orcid{0000-0003-0884-1499}

\thanks{This research has been supported by the MIUR under grant ``Dipartimenti di eccellenza 2018-2022'' of the Department of Computer Science of Sapienza University  and by the ``Sapienza information-based Technology InnovaTion Center for Health - STITCH''.}
\thanks{L. M. Author is with the Translational and Precision Medicine Department, Sapienza University of Rome, Rome, 00185 Italy (e-mail: lorenzo.madeddu@uniroma1.it).}
\thanks{G. G. Author is with the Translational and Precision Medicine Department, Sapienza University of Rome, Rome, 00185 Italy (e-mail: giorgio.grani@uniroma1.it).}
\thanks{P. V. Author is with the Computer Science Department, Sapienza University of Rome, Rome, 00185 Italy (e-mail: velardi@di.uniroma1.it).}
}

\maketitle

\begin{abstract}
\textit{Objective:} Human-curated disease ontologies are widely used for diagnostic evaluation, treatment and data comparisons over time, and clinical decision support. The classification principles underlying these ontologies are guided by the analysis of observable pathological similarities between disorders, often based on anatomical or histological principles. Although, thanks to recent advances in molecular biology, disease ontologies are slowly changing to integrate the etiological and genetic origins of diseases,   nosology still reflects this ``reductionist'' perspective. Proximity relationships of disease modules (hereafter DMs) in the human interactome network are now  increasingly  used in diagnostics,  to identify pathobiologically similar diseases and to support drug repurposing and discovery. On the other hand, similarity relations induced from structural proximity of DMs also have  several limitations, such as incomplete knowledge of disease-gene relationships  and reliability of clinical trials to assess  their validity.  The purpose of the study described in this paper is to shed more light on disease similarities by  analyzing the  relationship  between categorical proximity of diseases in human-curated ontologies and structural proximity of the related DM in the interactome.
\textit{Method:} We propose a methodology (and related algorithms) to automatically induce a    hierarchical structure from proximity relations between DMs, and to  compare this structure with a human-curated disease taxonomy. 
\textit{Results:}  We demonstrate that the proposed methodology allows to systematically analyze  commonalities and differences among structural and categorical  similarity of human diseases,  help refine and extend human disease classification  systems, and identify promising network areas where new disease-gene interactions can be discovered.

\end{abstract}

\vspace{-9pt}
\section{Introduction}
\label{sec:introduction}
Network medicine is a new paradigm of medicine that applies network science and systems biology approaches to study diseases as a consequence of pathobiological processes that interact in a complex network.  Pieces of evidence in this field show that if a gene or molecule is involved in a disease, its direct interactors might also be suspected to play some role in the same pathological process. According to this hypothesis, proteins involved in the same disease show a high propensity to interact with each other, a property referred to as \textit{ ``disease module hypothesis''} \cite{networkmedicine, menche2015uncovering}. This  hypothesis suggests that, if we identify a few disease components, the other disease-related components are likely to be located in their network-based ``neighbourhood'', called  \textit{disease module}. Under a biological perspective, ``a disease module represents a group of
network components that together contribute to a cellular function whose disruption results
in a particular disease phenotype'' \cite{networkmedicine}. 
Furthermore, subsequent studies have shown the tendency for biologically similar diseases to have their respective  modules located in adjacent or overlapping areas of the interactome \cite{menche2015uncovering, goh2007human, booknetwork}. 
According to Loscalzo et al. \cite{booknetwork} and Menche et al. \cite{menche2015uncovering}, ``proximity and degree of overlap of two disease modules (in the human interactome) has been found to be highly predictive of the pathobiological similarity of the corresponding diseases'' and ``network-based location of each disease module determines its pathobiological relationship to other diseases''. Indeed, different disease modules can overlap, so that perturbations caused by one disease can affect other disease modules that could lead to co-morbidity and pathogenetic mechanisms \cite{networkmedicine}. Analysing  interconnections within  disease modules can help reveal new disease genes, disease pathways, and identify possible drug targets or biomarkers for drug development and drug repurposing \cite{cheng2019network, networkmedicine}.
The tendency of phenotypically similar diseases to be close  or to overlap in the interactome
suggests the possibility of \textit{inducing a hierarchical and possibly categorical structure of  disease modules}, with specific and yet unexplored
relationships with existing disease classification taxonomies. 
Disease taxonomies play a key role in defining the mechanisms of human diseases, potentially impacting both diagnosis and treatment. However, as remarked in \cite{networkmedicine, loscalzo2007human,zhou2018systems}, contemporary approaches to the classification of human diseases are mainly based on 
anatomical pathological data and clinical knowledge.
Yet, modern molecular diagnostic tools have shown the shortcomings of this methodology, reflecting both a lack of sensitivity in identifying pre-clinical diseases and a lack of specificity in defining diseases unequivocally. On the other hand,  inducing disease relationships solely from disease modules in the interactome is hindered by incomplete knowledge of disease-related genes \cite{networkmedicine, omim}. 

In this study we propose a methodology  to systematically compare  categorical relationships automatically induced from proximity of disease modules in the human interactome  network, with manually crafted categories in human-curated  ontologies.   Detected commonalities and differences may suggest latent and unknown molecular properties of diseases,  help  improve the current understanding of the disease mechanisms,  and
facilitate precise clinical diagnosis consistent with molecular network properties \cite{Ghiassian2015, madeddu2020feature, cheng2019network, van2018gene}.

\vspace{-5pt}
\section{AIMS AND METHODS}
Network-based analyses of gene interaction data have helped to identify modules of disease-associated genes, hereafter referred to as disease modules (DMs), widely used to obtain both a systems level and a molecular understanding of disease mechanisms \cite{gustafsson2014modules}. Disease modules have been successfully used, for example,  to prioritize diagnostic markers or therapeutic candidate genes, and in drug repurposing \cite{cheng2019network, networkmedicine, ay2007drug}. 
However, according to Barabási et al.  \cite{networkmedicine}, these results have marginally influenced the disease taxonomies and, conversely, to the best of our knowledge, disease taxonomies have not been used to analyse disease modules. In this study we aim for the first time to integrate taxonomic and network-based disease categorization principles, with the following innovative contributions: 

\begin{enumerate}
\item to automatically induce a full-fledged hierarchical structure from proximity relations between DMs in the human interactome; \item to compare this structure  with a human-defined disease taxonomy (such as the Disease Ontology\footnote{\url{https://disease-ontology.org}}));  
\item to systematically identify categorical analogies and discrepancies between molecular and human-defined taxonomies. 
\end{enumerate}
Our research hypothesis is that  a   study of the

relationships between molecular-based and human-curated disease taxonomies could help  refine our knowledge on human diseases and
identify limitations and perspectives of current module-based computational approaches to the study of diseases. 
As shown in the experimental Section \ref{results}, our study has some possibly relevant clinical implications:

\begin{enumerate}
    \item To identify promising regions of the human interactome where new disease-gene relationships could be discovered\footnote{either exploiting data-driven methods or clinical experiments};
    \item To identify unexplored  molecular relationships among diseases;
    \item To extend, correct and refine human-curated taxonomies.
\end{enumerate}

\begin{figure}[ht!]
\centering
\includegraphics[width=1.\linewidth]{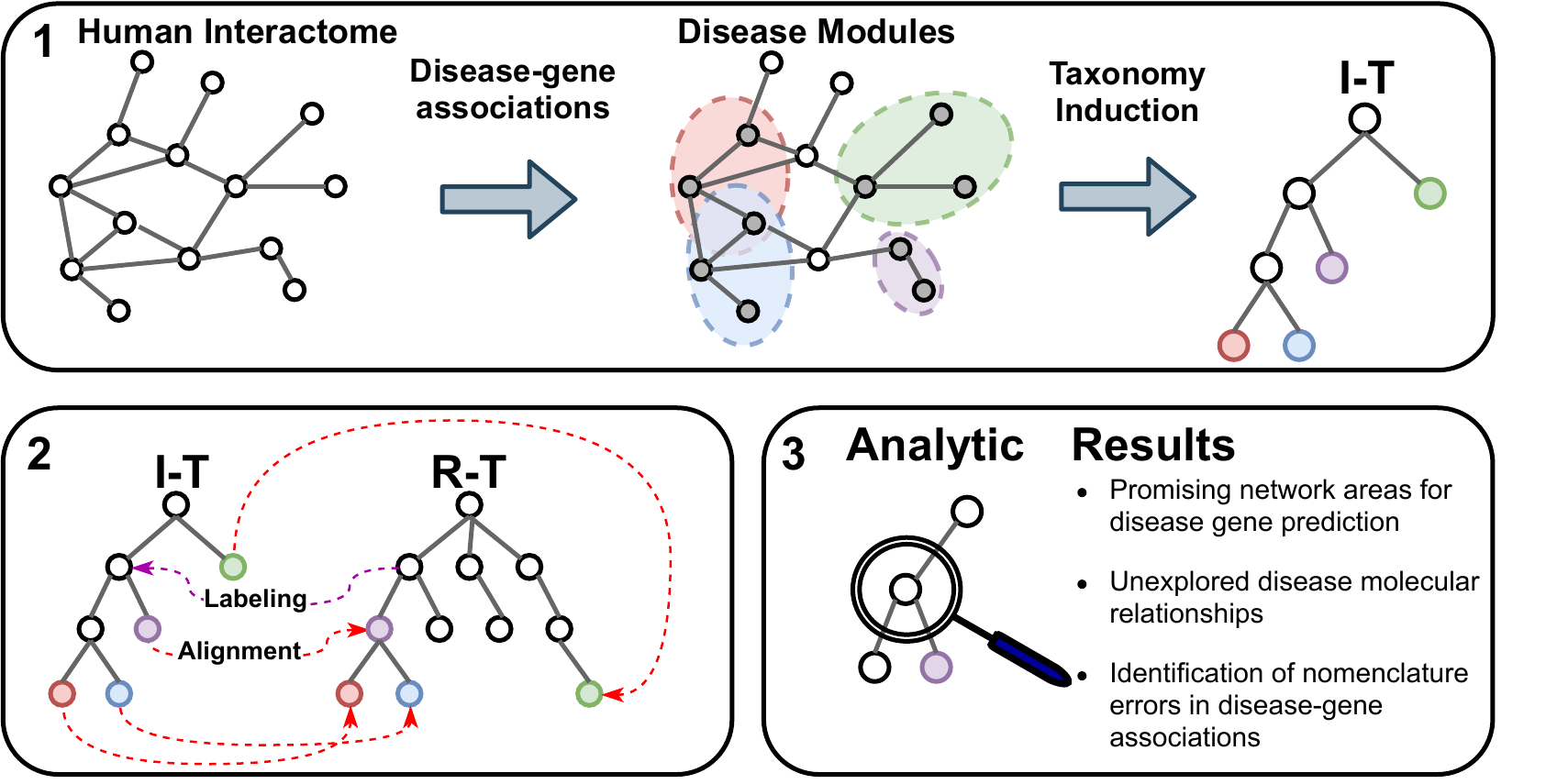}
\caption{The work-flow of our study. Box $1$ shows the taxonomy induction phase, Box $2$ represents the phase of taxonomy  alignment and labeling and Box $3$ summarizes the results of the analytic phase.}
\label{fig: workflow}
\end{figure}

\noindent
Figure \ref{fig: workflow} shows the workflow of the proposed approach, described in detail in the next Sections. The main phases are the following: \\
\noindent    
{\bf 1. Induction of a Taxonomy of Disease Modules:} 
    First, we automatically induce a hierarchical structure of diseases based on proximity relations of DMs in the human interactome. This taxonomic structure is hereafter referred to as the Interactome Taxonomy (I-T).\\
     \noindent
{\bf 2. Taxonomy alignment and labeling:} Next, we align the I-T taxonomy with a human-curated  \textit{reference  ontology} (hereafter R-T), by creating a mapping between disease nodes in both taxonomies (red arrows in Box 2 of Figure \ref{fig: workflow}). Finally,  a labeling algorithm finds the best map between categorical nodes in the R-T  and the unlabeled inner nodes of the  I-T (purple arrow in Box 2 of Figure \ref{fig: workflow}). \\
   \noindent
    {\bf 3. Systematic Comparative Analysis of I-T and R-T taxonomies:} The alignment between  I-T and  R-T supports a large-scale analysis of a vast collection of diseases  jointly from an ontological and molecular perspective.  
    We provide insights to refine state of the art nosology and knowledge on  disease interactions, by using our methodology to investigate the efficacy of the anatomical disease classification principle at the molecular level, identify nomenclature errors in disease-gene associations and discover  unexplored molecular mechanisms among diseases.

\vspace{-15pt}
\section{Construction of the Interactome Taxonomy (I-T)}
\label{sec:induction of I-T}
We induce a disease   taxonomy (named Interactome Taxonomy, I-T)  by applying hierarchical agglomerative clustering  to the human interactome network, exploiting proximity relations of disease modules. 
Hierarchical agglomerative clustering (HAC) is a set of greedy approaches that create a  hierarchy of clusters from unlabeled input data \cite{murtagh2017algorithms}. 
Given a distance matrix of seed clusters, the HAC algorithm iteratively merges  two clusters based on a selected  inter-cluster distance measure. Common methods to merge clusters are Average and Complete linkage \cite{murtagh2017algorithms}.
In our context, clusters are DMs in the human interactome network.  
However, due to the  high incompleteness of the disease-gene associations modeled in the human interactome \cite{Venkatesan}, disease modules are not molecularly well-defined and devoid of a clearly dense network-structure \cite{menche2015uncovering}. To cope with this problem, we use two alternative definitions of modules adopted in network science, which have been   commonly used in Network Medicine literature to physically identify disease modules \cite{menche2015uncovering, agrawal2018large}.\\ 
Given the human interactome network $G=(V, E)$ and a disease $d$ in a set of diseases $D_{it}$:
\begin{enumerate}
    \item \textbf {Induced Module:} The Induced Module $I_{d} = (V_{d}, E_{d})$ is a subgraph of $G$, where  $V_{d} \subseteq V$ is  the set of genes-nodes associated with $d$ and  $E_{d}$ is the set of gene-gene interactions $E_{d}=\{(u, v)|(u, v) \in E\;and\;u, v \in V_{d}\}$\cite{menche2015uncovering}.  This  definition includes in a DM all the disease genes but, due to the incompleteness of the network, it is usually a  graph with many connected components lacking a strong local structure \cite{agrawal2018large}.
    
    \item \textbf {Largest Connected Component (LCC) Module:} The  $LCC_{d}$ module is the largest connected component of $I_{d}$ \cite{menche2015uncovering, agrawal2018large}.  Unlike for the induced module $I_{d}$, $LCC_{d}$ usually has a denser local structure but may not include all the disease-related nodes $d$.
    \end{enumerate}
 
Given the human interactome $G$, a set of diseases $D_{it}$ and their disease modules $DM_{it}$ in $G$, hierarchical clustering is performed using a distance matrix of  disease modules  (defined as previously explained), based on the following network-based distance measure (Eq.\eqref{eq:dist})  (used, e.g., in \cite{menche2015uncovering, booknetwork, cheng2018network}):

\begin{footnotesize}
\begin{equation}
dist(A,B) = \frac{\sum_{a \in A}min_{b \in B}SP(a,b) +  \sum_{b \in B}min_{a \in A}SP(a,b)}{|A|+|B|}
\label{eq:dist}
\end{equation}
\end{footnotesize}

\noindent where A, B are respectively the set of nodes in modules $DM_A$ and $DM_B$ associated to diseases $d_{A}, d_{B} \in D_{it}$ and $SP$ is the shortest path length between two given nodes in  $G$.

In our experiments, we used two DM definitions (Induced Module and LCC)  and two cluster-merge methods (Average and Complete). The best solution among the four resulting I-Ts alternatives is identified using the methodology described in  Section \ref{sec:selection}.
 
\section{Taxonomy Alignment and labeling}
\label{sec:alignmentlabelling}
The result of the hierarchical clustering algorithm is a binary tree taxonomy, hereafter referred to as Interactome Taxonomy (I-T). I-T is a connected directed acyclic graph $T(V_{T}, E_{T})$ in which nodes $V_{T}$ represent disease concepts and edges represent ``is-a'' semantic relationships\footnote{Edge $(v, u)$ with $\;u, v \in V_{T}$ means that $v$ \textit{is a} kind of $u$.}. In our context, leaf-nodes (i.e. nodes with out-degree equal to zero) are ``specific'' diseases $D_{it}$, physically represented by the corresponding modules $DM_{it}$, while  inner nodes (i.e. non-leaf nodes) are disease categories $DC_{it}$.

Note that inner nodes $c \in DC_{it}$ are \textit{unlabeled}, and extensionally defined by the set of their subsumed disease nodes, 
referred to as the \textit{clusters} $C_c$ of nodes $c$. 
Similarly, given a ``reference'' human-crafted taxonomy, denoted as R-T, let $T(V_{R}, E_{R})$  be the set of its nodes and edges, $D_{rt}$  its disease (leaf) nodes, $DC_{rt}$ its  categorical (inner) nodes and $C_{c'}$  the clusters associated with categorical nodes $c' \in DC_{rt}$. Contrary to the I-T,  inner nodes in the R-T have also a human-defined label, the \textit{category name}.
\vspace{-9pt}
\subsection{Taxonomy Alignment}
\label{sec:alignment algorithms}
Whatever the choice of the R-T, the R-T and the I-T are expected to be defined on different sets of diseases nomenclatures, $D_{rt}$ and $D_{it}$. Furthermore, they are also expected to be structurally diverse. For example, R-T has usually a polyhierarchical structure, while I-T is by construction a binary tree.

To compare I-T and R-T we first need to create a mapping $M$  from
 $D_{it}$ to $D_{rt}$ nomenclatures, and next, to prune the hierarchies so that they include the same set of leaf disease nodes, a process that we call  \textit{taxonomy alignment}.

\begin{figure}[ht!]
\centering
\includegraphics[width=1.\linewidth]{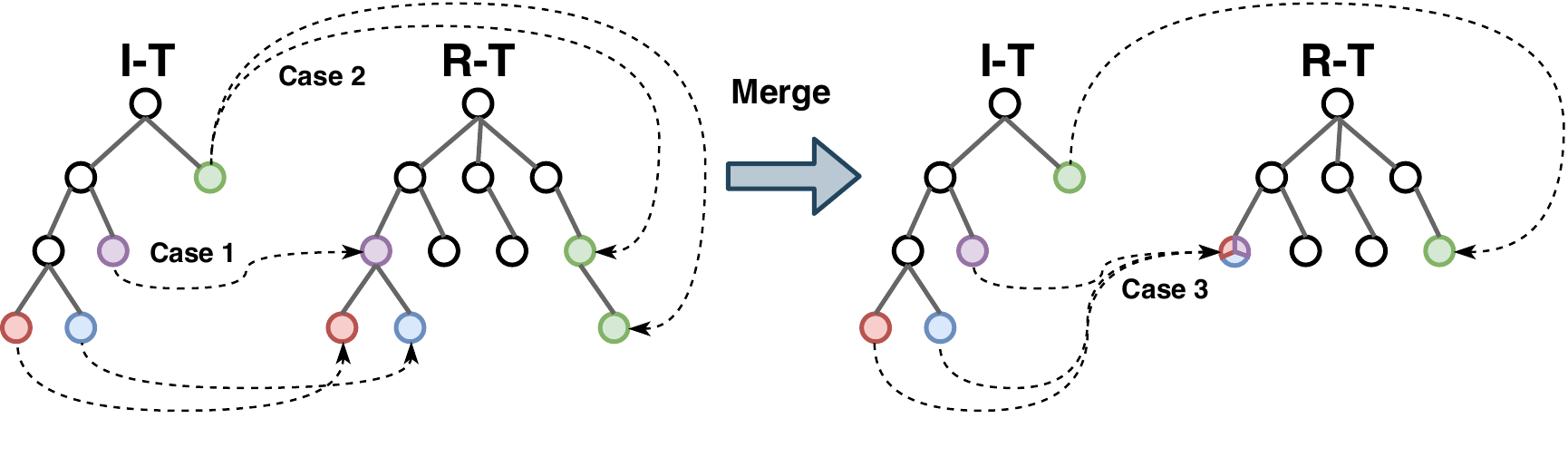}
\caption{Visual example of the \textit{merge} algorithm.}
\label{fig:merge}
\end{figure}

  Let $M$ be an available mapping of disease nomenclatures (see Section \ref{data} for details).  In  Figure \ref{fig:merge}, mappings among nodes of the two taxonomies are highlighted using the same colours. As shown,  $M$ is usually not one-to-one and we identify four cases:

\begin{enumerate}
    \item Case 1: Some leaf nodes $D_{it}$ in I-T may map onto inner nodes in the R-T.
    
    \item Case 2: Some  $D_{it}$  nodes may map onto multiple nodes in the R-T.
    
    \item Case 3: Some  $D_{it}$ nodes may map onto the same node in the R-T.
    
    \item Case 4: Some  $D_{it}$ nodes may have no mappings in the R-T \textit{and vice-versa}. These  nodes without mappings are depicted in Figure \ref{fig:merge} as  white leaf nodes in both taxonomies. 
    
\end{enumerate}

Our taxonomy alignment procedure consists of three algorithms: \textit{merge}, \textit{split}, and \textit{prune}. We apply \textit{merge} and \textit{split} to the R-T to solve  cases 1, 2 and 3; instead, \textit{prune} is applied to both the R-T and the I-T, to solve  case 4.  
\begin{algorithm}
\begin{footnotesize}
\SetAlgoLined
 \SetKwFunction{FMg}{Merge}

    \SetKwProg{Fn}{Function}{:}{}
      \Fn{\FMg{rt, mapping}}{
        diseases = keys(mapping)\;
        
        rtmapped = {}
        newMapping = Hashmap()
        \For{diseaseId in diseases:}
        {
            rtmapped.union(mapping[diseaseId])\;
            newMapping[diseaseId] = []\;
        }

        ancestors = rt.getColouredAncestors(rtmapped) \tcp*{Get the R\-T's coloured nodes with no R-T's coloured node among their ancestors}
        
        \For{ancestor in ancestors:}
        {
            diseasesDesc = rt.getMappingDescendant(ancestors, mapping)\tcp*{Get the diseases linked to R\-T's nodes descendant of the coloured ancestors}
            
            \For{diseaseId in diseasesDesc:}
            {
                newMapping[diseaseId].append(node)\;
            }
        }
        
        \For{ancestor in ancestors:}
        {
            descs = rt.getDescendants(ancestor)\;
            rt.removeNodes(descs)\tcp*{Turn every ancestor into a leaf}
        }
        
        mapping = newMapping\;
        \KwRet rt, mapping\;
  }
\label{algo:merge}
\end{footnotesize}
\caption{Merge}
\end{algorithm}
The \textit{merge} algorithm (see Algorithm  
1) turns in leaves all the R-T coloured inner nodes. If the node has non-coloured descendants (that is, its descendands do not map onto any disease module in I-T), these descendants are  removed. Else, the node and its coloured descendants are aggregated into a single multi-coloured node, as shown in Figure  \ref{fig:merge} (right). 

\begin{figure}[ht!]
\centering
\includegraphics[width=1.\linewidth]{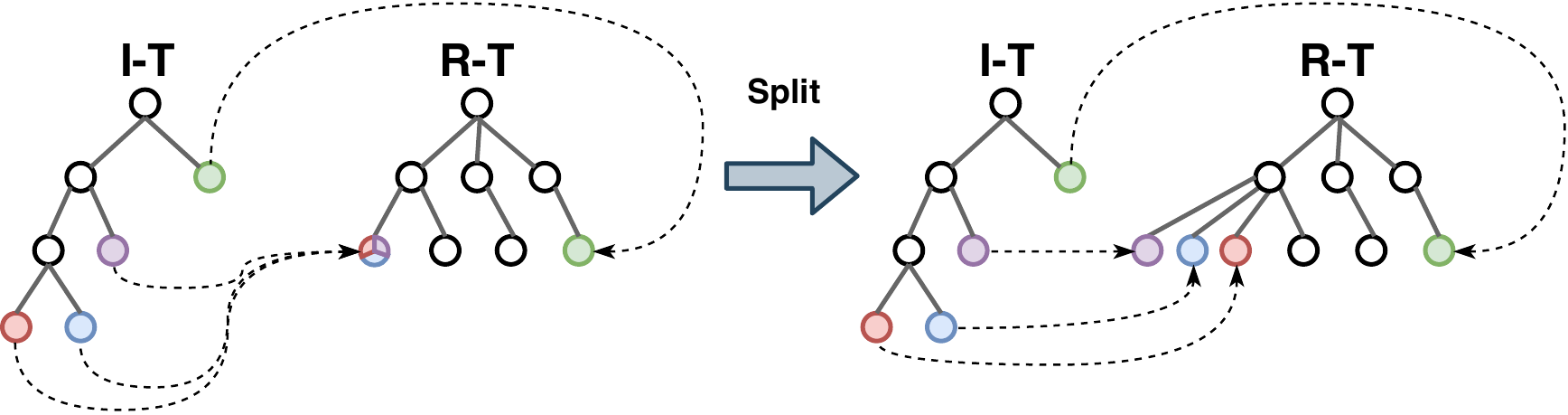}
\caption{Visual example of the  \textit{split} algorithm.}
\label{fig:split}
\end{figure}

\begin{figure}[ht!]
\centering
\includegraphics[width=1.\linewidth]{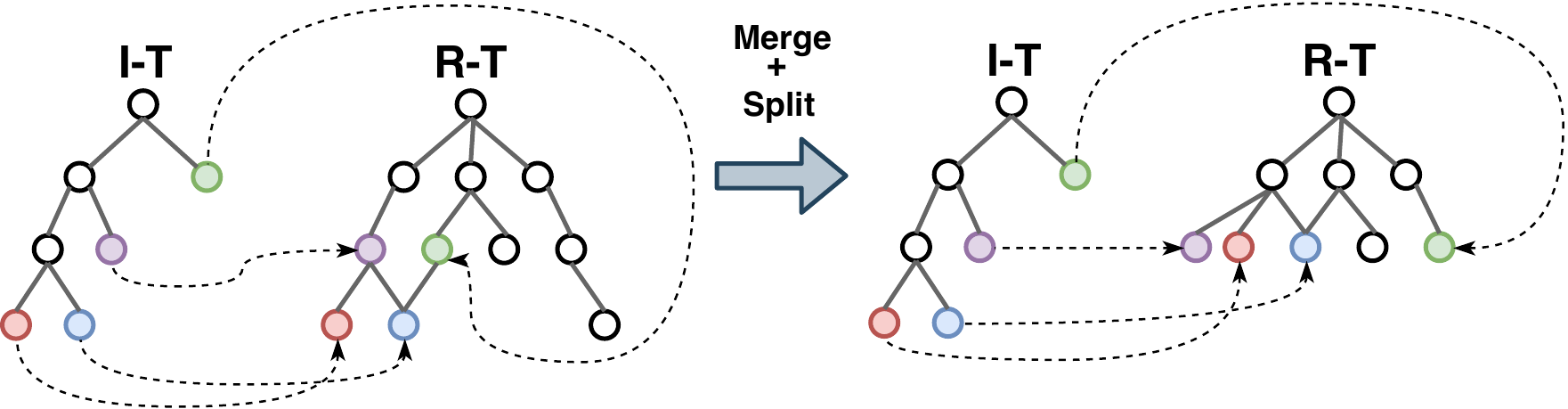}
\caption{Visual example of the \textit{merge} and  \textit{split} algorithms for a polyhierarchical case.}
\label{fig:merge-split}
\end{figure}

After the merge, the \textit{split} algorithm (see Algorithm 2) splits all the nodes with multiple colours (see Figure \ref{fig:split}). Note that these  nodes inherit the ancestors of the splitted multi-coloured node.  As a result of merge and split,  all coloured R-T nodes are now leaf nodes, and every I-T node points to its correspondent R-T node. Furthermore,  polyhierarchy in the R-T is preserved, as shown in  Figure \ref{fig:merge-split}.

\begin{algorithm}
\begin{footnotesize}
\SetAlgoLined
 \SetKwFunction{FSp}{Split}

    \SetKwProg{Fn}{Function}{:}{}
      \Fn{\FSp{rt, mapping}}{
        diseases = keys(mapping)\;

        \For{diseaseId in diseases:}
        {
            nodes = mapping[diseaseId]\tcp*{The R\-T's nodes linked to diseaseId}
            parents = rt.getParents(nodes)\tcp*{Get the set of parents of the given set of nodes}
            rt.addNode(diseaseId, parents)\;
            mapping[diseaseId] = diseaseId\;
        }
        
        \KwRet rt\;
  }
\label{algo:split}
\end{footnotesize}
\caption{Split}
\end{algorithm}

\begin{figure}[ht!]
\centering
\includegraphics[width=1.\linewidth]{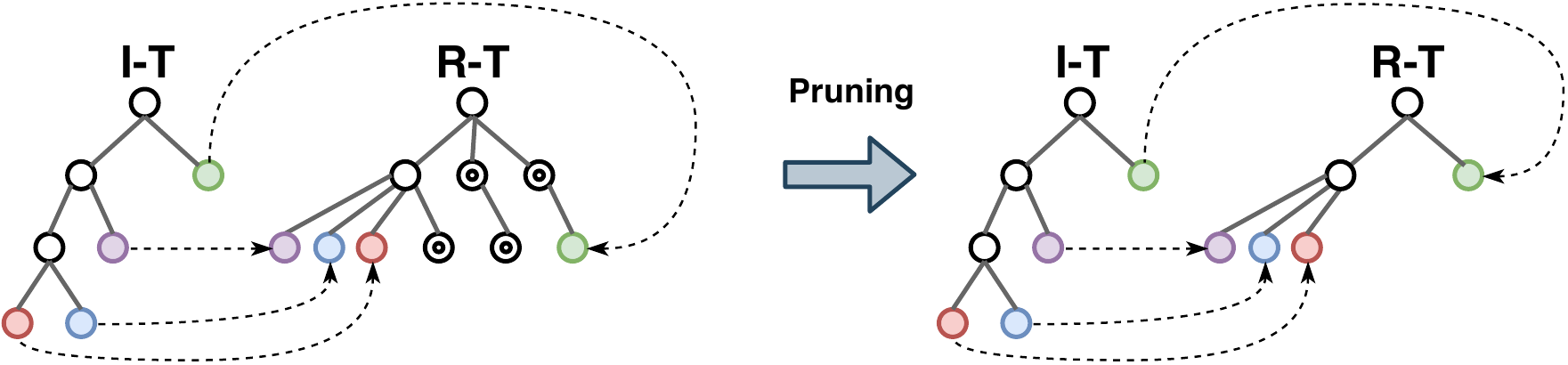}
\caption{Visual example of the \textit{prune} algorithm.}
\label{fig:pruning}
\end{figure}

Finally, the \textit{prune} algorithm (see  Fig. 
\ref{fig:pruning} and Algorithm  
3)  prunes both the R-T and the I-Ts by recursively removing survived  leaf nodes not linked by any mapping relation in $M$. These are shown with a double circle in Figure \ref{fig:pruning}.
As a final result, the R-T and the I-T have as leaf nodes the same set of  diseases, denoted as $D_{\cap}$.

\begin{algorithm}
\begin{footnotesize}
\SetAlgoLined
 \SetKwFunction{FPr}{Prune}

    \SetKwProg{Fn}{Function}{:}{}
      \Fn{\FPr{taxa, diseaseID}}{
        drop = taxa.leaves() - diseaseID\;
        
        \For{leaf in drop:}
        {
            taxa.remove(leaf)\tcp*{Remove the leaf from the taxonomy}
        }
        
        \For{leaf in taxa.leaves():}
        {
            \If{|rt.parents(leaf)| == 1 and rt.parents(leaf)[0].outDegree() == 1:}
            {
                taxa.replace(leaf)\tcp*{Recursively substitute the leaf with its parent until the leaf get  a parent with out-degree > 1 or more than one parent}
            }
        }
        \KwRet taxa\;
  }
\label{algo:prune}
\end{footnotesize}
\caption{Prune}
\end{algorithm}
\vspace{-9pt}
\subsection{Selecting alternative Induced Taxonomies}
\label{sec:selection}
As remarked in Section \ref{sec:induction of I-T}, the I-T is built using different definitions of disease modules and different inter-cluster similarity functions during agglomerative clustering. 
In this Section, we present a method to select the ``best'' I-T, among four I-Ts\footnote{Four aligned I-Ts resulting by the combination of the Induced and LCC disease module definitions with the Average and Complete clustering methods}, based on its structural and semantic similarity with the R-T. Note that a similarity function between two taxonomies can be computed only if they have been aligned. \\
Our method to compute the similarity between two taxonomies is based on the Lin semantic similarity  \cite{lin1998information}:

\begin{footnotesize}
\begin{equation}
 S_{T}(a, b):= \frac{2*IC_{T}(LCS_{T}(a, b))}{IC_{T}(a) + IC_{T}(b)}\label{eq:lin}
\end{equation}
\end{footnotesize}
\noindent 
where $IC$ is: 
\begin{footnotesize}
\begin{equation}
 IC_{T}(x):= -log(\frac{|leaves_{T}(x)|+1}{MaxLeaves_{T} + 1})
\end{equation}
\end{footnotesize} 

\noindent
where $a, b$ are two leaf nodes in a taxonomy $T$, and $LCS_{T}(a,b)$ is the least common subsumer of $a$ and $b$ in  $T$; $leaves_{T}(x)$ is the set of leaves descendant of $x$ and $MaxLeaves_{T}$ is the number of leaves in $T$.\\
The  Lin similarity increases when two  nodes are structurally close in a taxonomy, and decreases otherwise. Furthermore, by construction, the distance between two nodes is normalized with respect of the maximum distance, a property that is particularly useful when extending this measure to   compare taxonomies  with different depths.  This is a desirable property since the I-T is a binary tree and has a much higher depth than the R-T.  

To compare each of the four induced I-Ts with a selected R-T, first, we calculate the pairwise Lin similarity $S_{T}(d_i, d_j)$ 
 for each taxonomy $T$, where $\{(d_i, d_j)| d_i, d_j \in D_{\cap}\;and\;d_i \neq d_j \}$. Next, for each taxonomy $T$, we construct  a vector $v_T$ of $S_{T}(d_i, d_j)$  similarity values. Finally, we   calculate the cosine similarity between each of the induced taxonomies vectors $v_{IT_k} (k=1\dots 4)$  and $v_{RT}$. The intuition  is that, if  two taxonomies are similar, disease pairs that are ``close'' in one taxonomy should be ``close'' also in the other taxonomy, and those who are far apart in one taxonomy, should be far apart also in the other.  The ``best'' I-T is selected according to:\\
 \noindent
 $argmax_{k}(cosSim(v_{IT_k},v_{RT}))$.

The experimental application of this methodology is described in Section \ref{results}.

\vspace{-9pt}
\subsection{Semantic Labeling of the Interactome Hierarchy (I-T)}
\label{sec:labelling}
As previously noted, the inner nodes of the aligned I-T have no semantic labels. To facilitate a comparative analysis of I-T and R-T, we defined an algorithm to label each inner node in the  I-T with the most similar category label in the  R-T. In order to find the most similar R-T category node, we exploit the notion of  \textit{cluster} $C_c$ associated with a category node $c$ in a taxonomy, defined as the set of all its descendant disease nodes  that are also in $D_{\cap}$. Then, a \textit{labeling} algorithm (see Algorithm
4) labels every I-T disease category $c$ with the name of the R-T  category  $c'$ with highest  similarity score $sim(C_c,C_{c'})$  between the clusters of $c$ and $c'$\footnote{Note that the labeling method uses the full set of R-T categories to obtain more fine-grained labels
but the node clusters $C_c$ are defined on the common  disease set $D_{\cap}$ of the aligned taxonomies.} . To compute the similarity between two clusters, we use the  Jaccard coefficient, a popular measure of set similarity. 

\begin{algorithm}
\begin{footnotesize}
\SetAlgoLined
 \SetKwFunction{FMap}{labeling}

    \SetKwProg{Fn}{Function}{:}{}
      \Fn{\FMap{it, rt}}{
        categories = it.nodes() - it.leaves()
        labels = empty dictionary\;
        \For{category in categories:}
        {
            clusterIT = getCluster(it, category)
            catRT = getLabel(clusterIT, rt)\tcp*{Get the most similar R-T category}
            labels[category] = catRT\;
        }
        \KwRet labels\;
  }
\end{footnotesize}
\label{algo:label}
\caption{Labeling}
\end{algorithm}

\vspace{-15pt}
\section{Experimental set up}
\label{data}
This Section describes the data  sources used in our experiments.
To conduct a disease module analysis, we considered the most recent  release of the human protein-protein interaction network published by Barabasi et al. \cite{cheng2019network}, which is an extension of a highly cited and popular interactome used by Menche et al. \cite{menche2015uncovering}. The  network has $|V| = 16\,677$ proteins and $|E| = 243\,603$ physical undirected protein interactions. 

To construct disease modules, we collected disease-gene associations from DisGeNET \cite{pinero2020disgenet} with a GDA\footnote{GDA is a ``reliability'' score, for details see \url{www.disgenet.org/dbinfo\#section43}} score greater or equal of 0.3. Finally, we selected as disease modules the $948$ diseases with a set of disease genes of size at least $10$\footnote{smaller modules imply a limited knowledge of the related disease-gene associations to date, and may lead us to unreliable results.}.

We selected the Disease Ontology (DO) as Reference Taxonomy (R-T) \cite{schriml2012disease}. An alternative widely used reference ontology is ICD-9 (used, for example, in a work by Zhou et al. \cite{zhou2018systems}). However,  ICD-9  was built to facilitate the statistical study of disease phenomena, and arranged according to epidemiological properties and anatomical site. 
Hence, ICD-9 does not represent a good categorical framework for integrating network-based disease properties. 
Instead, the Disease Ontology is a classification of human diseases organised by  etiological agents such as infectious agents, clinical genetics and cellular processes. Therefore, even though the “localist” (i.e., anatomic and disciplinary classification) view of diseases is still a guiding principle, the DO also integrates the molecular insights of disease phenotypes.

By parsing the DO ``obo'' file\footnote{\url{http://www.obofoundry.org/ontology/doid.html}}, we generated a directed acyclic network hierarchy of $10012$ diseases and disease categories, $10061$ edges and $12 $ levels. 
To create a mapping $M$ between the two different nomenclatures, we used partial mappings directly provided  in DisGeNET and in the DO, that we further extended with the support of  clinicians to cover all the 948 DMs. 

To begin, we applied the  method described in Section \ref{sec:labelling} to select the best induced I-T taxonomy, i.e., the one  with the highest similarity with the selected R-T (namely, the Disease Ontology). Table \ref{tbl:rank} shows the result of this comparison. Note that similarity values are compared against those obtained by a random shuffling the disease nodes. 
Based on the results of Table \ref{tbl:rank}, we select the I-T taxonomy obtained using Induced Modules to represent DMs, and the Average linkage method to merge clusters during hierarchical clustering.  This  induced  taxonomic structure shows a higher similarity with the Disease Ontology and therefore represents a good basis for our study. Note however that all the compared methods  produce taxonomies with a similarity value significantly higher than the random baseline.  The observed differences are mainly due to some structural differences and to the positioning of  outliers (isolated DMs in the interactome).

\begin{table}[h!]
\centering
\scalebox{0.8}{
\begin{tabular}{|c|c|c|c|c|c|c|c|c|c|c|c|c|c|c|c|}
\hline
        & \multicolumn{2}{c|}{{\bf Induced}}  & \multicolumn{2}{c|}{\bf LCC}\\ \hline
Measure & Complete (RD) & { Average} (RD)  & Complete (RD) & Average (RD)\\ \hline
Cos. Sim. (\%) &    43.59  (28.55)    &  {\bf 46.33}   (29.84)  &    39.94   (28.58)    &  43.7   (29.71)\\
\hline
\end{tabular}%
}

\caption{\begin{footnotesize}Comparison of Lin-similarity vectors of the aligned  R-T (the Disease Ontology) and  I-T  taxonomies obtained with four different variants of the proposed taxonomy induction  method. The variants were obtained adopting two different network-based definitions of DMs (Induced and LCC) and two different techniques to merge clusters (complete and average). The value in the round brackets represents the average of  values generated by ten random distributions  of leaf nodes, leaving the taxonomy structure unchanged.\end{footnotesize}}
\label{tbl:rank}

\end{table}

\vspace{-20pt}
\section{Results}
\label{results}

Our research hypothesis in this work is that jointly analyzing the structural proximity of disease modules in the human interactome network and the semantic proximity of corresponding diseases in human-cured taxonomies could help both refine the classification of human diseases and
identify the limitations and perspectives of current module-based computational approaches to the study of diseases. In this Section, we summarize the major outcomes of a clinical analysis supported by the methodology presented in previous Sections. Our analysis is based both on the study of matching and unmatching pairs of R-T and I-T categories.

\vspace{-10pt}
\noindent
\subsection{Finding disease categories with a corresponding dense neighbourhood in the interactome. }
\label{dense neighbourhood}
First, we  conducted an analysis to reveal in the human interactome large neighbourhoods of disease modules associated with disease categories in R-T. Dense neighborhoods of diseases in the interactome network are useful to identify promising disease categories for disease gene prediction, drug repurposing and comorbidities detection.
To find these large neighbourhoods, we verified the existence of  topmost disease categories of the DO (our selected R-T) with a high overlapping with some inner (categorical) nodes in the  I-T. A  DO disease category $c'$ that is  ``well-represented'' by  an I-T category $c$ implies 
 a strong molecular proximity relationship among the  diseases in cluster $C_c$. Symmetrically, this implies  that there exists  a molecular mechanism that strengthens the classification principle of the DO category. 

We considered the 8 disease categories in the first level of the DO as the most general disease categories. 
To evaluate the degree of similarity between these DO categories and their most similar correspondents in the  I-T, we used the Jaccard similarity, i.e. the ``label score'' computed by the labeling algorithm of Section \ref{sec:labelling}. We also calculated the statistical significance of our results by computing the p-value over a random distribution. 
Table \ref{tbl:macro} provides an overview of the topmost DO  disease categories and their similarity degree with correspondent I-T categories induced from DM molecular network-proximity. In particular, we found that the DO disease categories that show a higher localization in a network neighborhood are ``disease of cellular proliferation'' and ``genetic disease''. This  means that tumors and genetic diseases are highly localized in two neighbourhoods of the human interactome. From a biological network perspective,  close DMs of ``disease of cellular proliferation''  are motivated by the fact that cancer diseases have similar genetic causes in differentiation and proliferation control genes such as the well-known $P53$ \cite{goh2007human, wood2007genomic,networkmedicine}. 
The second best matching category is  ``disease of anatomical entity'', i.e., disease grouped by human experts according to an anatomical  localization principle.  However, as shown in the Table \ref{tbl:macro}, the similarity value is high but not statistically significant. This is motivated by the fact that  diseases belonging to this topmost category are grouped in diverse sub-categories scattered over the network rather than in a large ``anatomical'' neighbourhood. 
To confirm this hypothesis, we performed a systematic automated pair-wise comparison among sub-categories of ``disease of anatomical entity''. We found that \textit{very rarely} category pairs belonging to different anatomical sub-systems  have overlapping clusters in the I-T, with some obvious and well documented exception, like nervous and respiratory systems, gastrointestinal and integumentary systems, musculoskeletal and cardiovascular systems \cite{chhabra2005cardiovascular, huang2012skin, maron2013hypertrophic}.  In other terms, our experiments show that the  validity of the anatomical classification principle is not  disproved by the DM localization hypothesis, at least,  given our state-of-the-art knowledge of disease-gene associations. This observation leads us to consider one limitation of the study presented in this Section, which stems from the high incompleteness of the human interactome \cite{Venkatesan}. It follows that, while positive results (disease categories corresponding to highly overlapping disease modules) are useful pieces of evidence to identify interesting areas of the interactome to discover new disease-gene associations, the absence of such evidence could be either motivated by the non existence of a similarity relation, or by a lack of knowledge on gene interactions in specific areas of the interactome.\\
\begin{table}[h!]
\centering
\scalebox{1.0}{
\begin{tabular}{|c|c|c|c|}
\hline
R-T (Disease Ontology) & Induced I-T\\\hline
\textbf{Disease Category Name} (size) & \textbf{Best Label Score} (P-value)\\\hline
disease of cellular proliferation  (255)  &  54.77\%  ($3.14\cdot10^{-20}$)\\\hline
disease of anatomical entity   (434)   &    50.05\%  ($0.08$)\\\hline
genetic disease  (12) &   41.66\%  ($6.14\cdot10^{-10}$)\\\hline
disease by infectious agent  (10) &  30\%  ($1.92\cdot10^{-7}$)\\\hline
physical disorder   (21) &   26.09\%  ($1.51\cdot10^{-9}$)\\\hline
disease of mental health  (76)   &   21.51\% ($1.06\cdot10^{-13}$)\\\hline
syndrome (42) &  21.27 \% ($8.69\cdot10^{-11}$)\\\hline
disease of metabolism   (55)  &   16.36\%  ($4.66\cdot10^{-11}$)\\\hline
\end{tabular}%
}
\caption{Correspondence among topmost DO categories and the induced taxonomy. 
}
\label{tbl:macro}
\end{table}

\vspace{-15pt}
\subsection{Finding unexplored structural relations between disease categories. }
\label{sec:unexpected}
A more interesting result would be to identify ``unexpected''  and unexplored neighborhoods in the I-T, e.g., disease  categories that are not presently connected in human-curated taxonomies but whose strong molecular similarities suggest that one such connection should be exploited to enrich the R-T ontology. 
To help finding these  relations we developed a visual tool to explore the   I-T in a more systematic way. Supported by this tool, clinical experts have identified, among the others, the following interesting results:
there exist strong unexpected molecular relationships between glaucoma and pulmonary arterial hypertension,
 cholestasis and chronic obstructive pulmonary diseases (COPD), peroxisomal diseases and ciliopathy-related syndromes. \textit{We remark that these categorical relationships can be detected in the I-T thanks to the labeling algorithm}. 
 
 By delving into these  relationships, we were able to find confirmations in very recent clinical studies. 
For example, Gupta et al. \cite{gupta2020secondary} and Lewczuk et al. \cite{lewczuk2019ocular} shed light on common molecular mechanisms and manifestations between pulmonary hypertension and glaucoma through multiple case reports. Instead, Tsechkovski et al.  \cite{tsechkovski1997a1} observed that cholestasis and COPD patho-mechanisms are mediated by common molecular components like the Alpha 1-antitrypsin protein. However, the relationship between Alpha 1-antitrypsin mutations and liver disease is debated and yet to be elucidated \cite{schneider2020liver}. It is interesting to note that there is an emerging hypothesis connecting gut, liver and lung as playing a key role in the pathogenesis of COPD  \cite{young2016gut}. Finally, Miyamoto et al.  Zaki et al. \cite{zaki2016pex6} found biological mechanisms between peroxisomal diseases and ciliopathy related syndromes (e.g. Joubert syndrome, Bardet-Biedl syndrome, Jeune syndrome).  
In conclusion, recent clinical evidence confirms that these detected relationships could be used to extend the DO.
Other unexplored strong relationships that we identified lack at the moment support from published studies\footnote{a clinical confirmation of our findings is clearly outside the scope of this research, although it represents a study hypothesis for further research by clinicians in the field.}, however the results reported above demonstrate the relevance and potentials of our proposed methodology.
\vspace{-9pt}

\subsection{Detection of nomenclature errors in disease-gene associations.}
\label{sec:nomenclature}

Finally,  we tried  to identify  ``unconvincing'' strong matches between R-T and I-T categories.
An in-depth analysis  by clinicians has led to the detection of a number of nomenclature errors in DisGeNET.

Disease modules in the interactome have been identified, as discussed in Section \ref{data}, using disease-gene associations in DisGeNET, one of the most widely used association databases. Publicly available disease-gene associations databases are manually or computationally curated and some of them integrate other disease-gene collections. However,  especially for ambiguous diseases with similar names, all these mechanisms are prone to nomenclature errors resulting in wrong  disease-gene associations.  Although in our work we selected only associations with a high GDA score,  errors might still survive.

The identification of wrong disease-gene associations is of primary importance both for disease gene discovery and clinical diagnoses. Indeed, on the one hand, disease gene discovery tools, using wrong disease-gene associations, would  make wrong predictions. On the other hand, clinicians usually make and justify diagnoses using the disease-gene associations contained in public databases (as we said, DisGeNET is one among the most widely used resources) leading to wrong diagnoses or therapies for a patient. 
Here, we demonstrate that our framework may  facilitate the detection of wrong disease-gene associations in public databases, caused by nomenclature errors. 
To this end, supported by clinicians, we identified a number of DO disease categories with an unconvincingly high overlapping with I-T  inner nodes. Specifically, a clinical expert  analyzed all DO/I-T  categories pairs with a  Jaccard similarity score greater or equal than $90\%$. Then, we manually verified the DisGeNET  pieces of evidence supporting the related disease-gene associations. 

Several nomenclature errors were found, among which we cite the following:
 ``hyper-IgM Immunodeficiency Syndrome'', ``obstructive lung disease''  and  ``bone  remodeling  disease'' have several wrong disease-associations. For example, the ``hyper-IgM immunodeficiency syndrome'' is divided into five types by  genetic association. In particular, ``hyper-IgM immunodeficiency syndrome'' type 2 is characterized by mutations of the \textit{AICDA} gene; type 3 by mutations of the \textit{CD40} gene, type 5 by mutations of the \textit{UNG} gene. However, DisGeNet links the three disease types  to the same three genes\footnote{\url{https://www.disgenet.org/browser/0/1/0/C1720956/}}. This error is probably due to a wrong integration of the disease-gene associations of the CTD database that does not characterize the ``hyper-IgM immunodeficiency syndrome'' in types\footnote{\url{http://ctdbase.org/detail.go?type=disease&acc=MESH\%3aD053306\#descendants}}.  
More in detail, in ``obstructive lung disease'' the pulmonary emphysema, focal emphysema, panacinar emphysema, centrilobular emphysema diseases have the same 12 disease-gene associations almost all supported by the same published study, but related only to the pulmonary emphysema or the generic category of emphysema. The same  happens in ``bone remodeling disease'' for the following diseases: osteoporosis, age-related osteoporosis, post-traumatic osteoporosis, senile osteoporosis.

\vspace{-9pt}
\section{Discussion}
\label{discussion}
We believe that the biomedical understanding of diseases is on the edge of a radical change. The disease module hypothesis, with its relevant applications to disease-gene discovery and drug repurposing, is leading the revolution of bio-medical research of the future. For these reasons, we deem it fundamental to discover the degree of correspondence between disease similarity relations induced from the proximity of their related disease modules, and categorical similarity in  human-curated disease taxonomies. 
We developed a methodology  to analyse relationships between diseases by leveraging, in a novel way, both taxonomic and molecular aspects. 
The proposed methodology supported a systematic analysis of human-crafted disease categories and their relationships with the DM molecular network-proximity. In particular, we found that some disease in ``disease of cellular proliferation'' and ``genetic disease'' form promising large disease network-neighbourhoods that could be exploited by network  analysis methods for disease-gene detection. Next, we evaluated the consistency of the ``disease anatomical entities'' at the molecular level and found that there is no  strong evidence of a network-neighbourhood of anatomical entities but, contrarily,  disease neighbourhoods related to anatomical systems are scattered. Furthermore, we used our methodology to find unexplored strong molecular relationships between ``specific'' disease categories, such as glaucoma and pulmonary hypertension,  diseases that are distant in human-crafted taxonomies but appear to be  related by co-morbidities and pathogenesis at the molecular level. Finally, we have been able to detect wrong disease-gene associations caused by nomenclature errors in public databases, that could potentially bias disease gene prediction methods and induce wrong clinical diagnoses.\\
\vspace{-9pt}
\section{Related Works}
\label{related works}
To the best of our knowledge, only one study has analysed in detail the molecular and categorical properties of DMs, as proposed in this project. 
Zhou et al. \cite{zhou2018systems}, as a response to the limits of the contemporary disease taxonomies, demonstrate the inconsistency of the ICD-9 diagnostic classification system with the disease relationships in molecular networks. They propose a New Classification of Diseases (referred to as \textit{NCD}) to capture the molecular diversity of diseases and define clearer boundaries in terms of both phenotypical similarity and molecular associations. 
The purpose of their study is to reform the ICD-9 by constructing a NCD in three steps. First, they create a network of diseases connected by molecular and phenotypic similarities. The disease nodes of the network define the low level of NCD, that is, the leaf nodes of the hierarchy. Then, the authors apply an overlapping community detection algorithm to the disease network to generate overlapping disease categories, representing the middle level of the NCD. Finally, they apply a non-overlapping community detection algorithm to the previously identifies network communities and generates disease categories for the top level of the NCD. 

As highlighted by Zhou et al. \cite{zhou2018systems}, NCD is a simplified three-level hierarchical structure without explicit mappings with the reference classification system, the ICD-9. We build on \cite{zhou2018systems}, by proposing  a method to create a full-fledged taxonomy with category labels, to favour a more systematic comparison between the automatically induced and manually created taxonomies.
Furthermore, as mentioned in Section \ref{data} the use of ICD-9 as a reference taxonomy has some drawbacks.
ICD-9  has been designed to promote international comparability in the classification and presentation of epidemiology and mortality statistics. Its hierarchical structure is not based on aetiology, but rather on anatomical and disciplinary principles,
to facilitate the statistical study of disease phenomena, and arranged according to epidemiological properties and anatomical site. 
Hence, ICD-9 does not represent a good categorical framework for integrating network-based disease properties. Instead, the Disease Ontology (DO) has the purpose of identifying “commonalities of diseases located in a common molecular location, originating from a particular cell type or resulting from a common genetic mechanism” \cite{schriml2012disease}. Therefore, even though the “localist” view of diseases is still a guiding principle, the DO also exploits the molecular insights of disease phenotypes, thus representing a more appropriate baseline ontology. \\
\vspace{-15pt}
\section{Conclusions}
\label{conclusions}

We presented a novel disease module analytic strategy leveraging both a molecular and  taxonomy perspective, providing new insights into the molecular mechanisms of diseases and a  way to refine  human-curated taxonomies. Our methodology has supported clinically relevant findings, such as promising areas of the interactome to discover new  disease gene associations, unexplored disease molecular relationships, and nomenclature errors in disease-gene databases.

One limitation of our study arises from the highly incomplete state of the art  knowledge on disease-related genes. This resulted in a limited mapping between human-crafted taxonomies and our induced hierarchy of disease modules (about 12\% of DO diseases), and furthermore prevented the interpretation of some evidence concerning unobserved molecular relationships, which could be either motivated by the non existence of such relations, or by the lack of knowledge on gene interactions in specific areas of the interactome.   

\bibliographystyle{ieeetr}
\bibliography{references}

\end{document}